\documentclass[12pt,preprint]{aastex}

\newcommand{\ep}{\epsilon}

\newcommand{\dpa}{\partial}

\shorttitle{Planet Migration in Turbulent Disks.}
\shortauthors{Uribe, Klahr, Flock and Henning}

\begin{document}

\title{3D MHD Simulations of Planet Migration in Turbulent Stratified Disks}

\author{A. L. Uribe, H. Klahr, M. Flock and Th. Henning}
\affil{Max-Planck-Institut f\"{u}r Astronomie,\\ Koenigstuhl 17, D-69117 Heidelberg, Germany.}
\email{uribe@mpia.de}

\begin{abstract}
We performed 3D MHD simulations of planet migration in stratified disks using the Godunov code PLUTO, where the disk is turbulent due to the magnetorotational instability. We study the migration for planets with different planet-star mass ratios $q=M_{p}/M_{s}$. In agreement with previous studies, for the low-mass planet cases ($q=5\times10^{-6}$ and $10^{-5}$), migration is dominated by random fluctuations in the torque. 
For a Jupiter-mass planet $(q=M_{p}/M_{s}=10^{-3}$ for $M_{s}=1M_{\odot})$, we find a reduction of the magnetic stress inside the orbit of the planet and around the gap region. After an initial stage where the torque on the planet is positive, it reverses and we recover migration rates similar to those found in disks where the turbulent viscosity is modelled by an $\alpha$ viscosity. For the intermediate-mass planets ($q=5\times10^{-5}, 10^{-4}$ and $2\times10^{-4}$) we find a new and so far unexpected behavior. In some cases they experience sustained and systematic outwards migration for the entire duration of the simulation. For this case, the horseshoe region is resolved and torques coming from the corotation region can remain unsaturated due to the stresses in the disk. These stresses are generated directly by the magnetic field. The magnitude of the horseshoe drag can overcome the negative Lindblad contribution when the local surface density profile is flat or increasing outwards, which we see in certain locations in our simulations due to the presence of a zonal flow. The intermediate-mass planet is migrating radially outwards in locations where there is a positive gradient of a pressure bump (zonal flow).
\end{abstract}


\section{INTRODUCTION}

Understanding why and how fast planets migrate is fundamental to explaining the observed distribution of exoplanets and constraining planet formation timescales and efficiencies \citep{2005A&A...434..343A}. The basic principle behind the migration of planets in protoplanetary disks is the transfer of angular momentum between the planet and its disk. This transport process occurs at Lindblad resonances and, in a locally isothermal disk, typically leads to fast inwards migration.  
\citep{1980ApJ...241..425G, 1984ApJ...285..818P,1986Icar...67..164W,2002ApJ...565.1257T}. 
This is the standard type I migration scenario which applies to low- to intermediate-mass planets where the specific torque is a linear function of the planet mass \citep{1997Icar..126..261W}.
\par
If the planet is massive enough (the mass depending on the viscosity and the pressure scale height), the tidal forces on the disk can eventually overcome the pressure gradient and the viscous transport, causing gap opening around the planet orbit. This migration regime, referred to as type II, in which the planet-disk interaction can no longer be described by linear perturbation theory, is then conceptually very different from the type I regime. Due to the gap opening, it is possible for the torques on the planet to cancel in such a way that the evolution of the planet's position is determined by the viscous transport of gas in the disk, making the planet move with the disk on viscous timescales \citep{1997Icar..126..261W,2003MNRAS.341..213B}. 
\par
Current numerical and analytical calculations estimate migration timescales to be a small fraction of the expected disk lifetime, which creates a problem for the survival of planetary cores. Gas planet cores need to reach a critical mass before the onset of runaway gas accretion \citep{2008IAUS..249..223I}. It is well established that planet population synthesis models together with giant planet formation models require a much less efficient type I migration to reproduce the observed distribution of exoplanets \citep{2008PhST..130a4022B,2009A&A...501.1161M,2005A&A...434..343A,2004ApJ...604..388I,2007arXiv0709.1375I,2002A&A...394..241T}. Nevertheless, core survival mechanisms have also been proposed to solve the timescale problem without resorting to an artificially reduced Type I migration rate \citep{2003MNRAS.341.1157T,2005MNRAS.363..943F,2006ApJ...644.1214T}.
 
Deviations from linear theory have been found in a number of three dimensional calculations. \citet{2002A&A...387..605M} and \citet{2003ApJ...586..540D} and later \citet{2006ApJ...652..730M} found that for intermediate-mass planets (around $M_{p}=1\times10^{-4}M_{s}$), the torques on the planet can be significantly lower and even reverse sign when the local surface density profile of the disk is flatter ($\Sigma \sim r^{\alpha}$ with $\alpha=0-0.5$) than in the usually assumed Minimum Mass Solar Nebula (MMSN) model $\Sigma \sim r^{-1.5}$. This is found for a certain range of the disk viscosity. In this case, the torques from the corotation region can become important. The fluid elements that are librating (moving in horseshoe orbits in the corotation region) orbit on a U-turn trajectory at trailing and leading sides of the planet. These fluid elements exert a torque on the planet at each U-turn, which is a symmetric effect on both sides of the planet in an inviscid disk; therefore there is no net torque coming from this region after a few librating periods. This is referred to as the "saturation" of the corotation torque. In the presence of viscosity, if the viscous crossing timescale across the horseshoe region of accreting disk material is smaller than the libration timescale, the torques exerted by fluid elements around the U-turn are not symmetric at each side of the planet, creating a net positive torque that can be sustained. This is refereed to as the "unsaturated" corotation torque, and it can depend on the surface density and on the width of the horseshoe region that in turn depends on the planet mass \citep{1992NYASA.675..314W}.

\par
So far most numerical studies of migration and/or gap formation have concentrated on the quasi-laminar disk case, where Navier-Stokes shear viscosity is included in order to model the viscous stresses resulting from including turbulence in the disk (e.g. \citet{2000MNRAS.318...18N, 2003MNRAS.341..213B, 1999ApJ...514..344B,2006MNRAS.370..529D,2000MNRAS.315..823P,2006Icar..181..587C} and many more). 
There has been strong interest in simulating planet-disk interactions in turbulent disks, where the turbulence is magnetically generated by the magneto-rotational instability (MRI) \citep{1991ApJ...376..214B,1998RvMP...70....1B}. 
Only ideal MHD has been considered so far in global simulations. The disk is assumed to be fully ionized and the magnetic diffusivity is negligible.  \citet{2003ApJ...589..543W} looked at gap formation by intermediate- and large-mass planets in turbulent unstratified disks and the local internal stresses around the planet. In the MHD case, they found the gap to be shallower and wider compared to the laminar HD case; the Maxwell stresses in the disk dropped in the vicinity of the planet's orbit. \citet{2003MNRAS.339..983P} performed a comprehensive study of protoplanets embedded in MHD-turbulent unstratified disks. They found that for low mass planets, Type I migration is no longer effective due to large fluctuations in the torque. No convergence was reached due to fluctuations of the torque on timescales longer than the orbital period and short simulation timescales. However, the torques for planets more massive than $30M_{\oplus}=0.1M_{Jup}$ were found to converge to the standard Type I migration torques after long-time averaging \citep{2003MNRAS.339..993N, 2004MNRAS.350..829P, 2004MNRAS.350..849N}. For low-mass protoplanets, \citet{2005A&A...443.1067N} studied the long-term evolution of the orbital elements and particularly the excitation of eccentricity by turbulent fluctuations. The evolution of the orbital elements of particles in MHD turbulence has also been studied using shearing unstratified boxes \citep{2009ApJ...707.1233Y} and stratified boxes including a dead zone \citep{2007ApJ...670..805O}. To avoid the expensive MHD simulations, other approaches have been taken, such as modeling the turbulence as a time and space varying forcing in a laminar disk model \citep{2004ApJ...608..489L}. In this case, depending on the amplitude of the forcing, type I migration can be overcome by the random fluctuations in the torque, and random walk motion will be superimposed on the smooth inward migration. \citet{2010ApJ...709..759B} used a similar turbulent forcing model and studied the unsaturation of the corotation torque due to turbulence. Depending on the amplitude of the turbulence, the corotation torque is found to be unsaturated to a certain level, making the total torque increase accordingly (become less negative), slowing down inwards migration. Other approaches include the analytical description of stochastic migration of low-mass planets using a diffusion-advection equation \citep{2006ApJ...647.1413J,2009ApJ...701.1381A} and coupling N-body simulations with a random forcing to study the accretion and formation of low-mass planets \citep{2007Icar..188..522O}. Recently, \citet{2010MNRAS.409..639N} examined the velocity dispersion of 1 m to 10 km planetesimals embedded in a turbulent disk, using 3D MHD simulations and neglecting stratification, and characterized the stochastic gravitational perturbations felt by planetesimals due to MHD turbulence.
\par
In this paper, we study planet migration in stratified 3D MHD-turbulent disks for planet masses in the Type I and II mass range. In Section \ref{setup} we describe the numerical setup of our simulations, and the initial conditions for the disk and the magnetic field, before the addition of a planet. In section \ref{results} we present the results of our simulations and finally in section \ref{conc} we discuss our results.

\section{SIMULATION SETUP}\label{setup}

Simulations where performed using the finite volume fluid dynamics code PLUTO \citep{2007ApJS..170..228M}.  In the code, time stepping is done using a second order Runge Kutta scheme, while the spatial integration is performed using linear interpolation through the second order TVD scheme. The Riemann fluxes are computed using the HLLC and HLLD solvers for the HD and MHD cases, respectively. The code uses the Constrained Transport method for preserving a divergence-free magnetic field \citep{2005JCoPh.205..509G}. The numerical setup for the MHD case follows the setup presented in
\citep{2010A&A...516A..26F}. The MHD equations in the isothermal approximation (no energy equation) are given by 
\begin{eqnarray}
\frac{\dpa \rho}{\dpa t} + \mathbf{\nabla}\cdot (\rho \mathbf{v}) = 0 \\
\frac{\dpa \mathbf{v}}{\dpa t} + (\mathbf{v}\cdot\mathbf{\nabla})\mathbf{v} + \frac{1}{\rho}\mathbf{B}\times(\mathbf{\nabla}\times\mathbf{B}) 
= -\frac{1}{\rho}\mathbf{\nabla} P - \mathbf{\nabla} \Phi_{g}\\
\frac{\dpa \mathbf{B}}{\dpa t} + \mathbf{B}(\mathbf{\nabla} \cdot \mathbf{v}) - (\mathbf{B}\cdot\mathbf{\nabla}) \mathbf{v} + (\mathbf{v}\cdot\mathbf{\nabla})\mathbf{B} = 0 
\end{eqnarray}
The potential $\Phi_{g}$ includes contributions from the star and the planet. We work in spherical coordinates $(r,\theta,\phi)$, where the computational domain is given by $r\in [1,10]$, $\theta \in [\pi/2 -0.3,\pi/2 +0.3]$ and $\phi \in [0,2\pi]$. The grid resolution is $(N_{r},N_{\theta},N_{\phi})=(256,128,256)$ and it is centered in the center of mass of the planet-star system. The boundary conditions for the velocities and magnetic field are periodic in the vertical ($\theta$ boundary) and azimuthal directions and reflective in the radial direction, except for the transverse magnetic field component, which reverses its sign at the radial boundary. Buffer zones are defined at the radial boundaries to avoid boundary effects, where for $1<r<2$ the magnetic resistivity is given by $\eta=2\times10^{-4}(2-r)$ and for $9<r<10$ the resistivity is $\eta=1\times10^{-4}(r-9)$.\\

\subsection{Disk and Planet Models}\label{sec:models}
As an initial condition we take a gas disk in sub-Keplerian rotation around a solar mass star. The azimuthal velocity is given by
\begin{equation}
v_{\phi}= \sqrt{v_{k}^{2} - c_{s}^{2}(a - 2b)},
\end{equation}
where $v_{k}$ is the Keplerian velocity and $a=3/2$ and $b=0.5$ are the exponents of the radial power law distribution of the density $\rho\propto r^{-a}$ and sound speed $c_{s}=c_{0}(r\sin\theta)^{-b}$. The initial density distribution is given by
\begin{equation}
\rho(r,\theta) = (r\sin\theta)^{-3/2}\exp\left(\frac{\sin\theta-1}{c_{0}^{2}}\right).
\label{eq:dens_prof}
\end{equation}
The disk is described by a locally isothermal equation of state $P=c_{s}^{2}\rho$. The ratio of the pressure scale height $h$ to the radial coordinate of the disk is taken to be a constant such that $h=H/(r\sin\theta)=0.07$.\\
The gravitational potential of the planet is given by a softened point-mass potential
\begin{equation}
\Phi_{p}(\mathbf{r}) = -\frac{GM}{(|\mathbf{r}-\mathbf{r}_{p}|^2 + \ep^{2})^{1/2}}
\end{equation}
where $\ep$ is the softening parameter, needed to avoid numerical divergence near the position of the planet and 
\begin{eqnarray}
|\mathbf{r}_{p}-\mathbf{r}|^2 &=& r_{p}^{2} + r^{2} -2r_{p}r(\sin\theta_{p}\sin\theta  \nonumber \\
&\times&\cos(\phi_{p}-\phi) + \cos\theta_{p}\cos\theta)
\end{eqnarray}
is the distance between the planet and a gas particle in the disk. For all the simulations $\ep$ is set to be a fraction of the Hill radius $\ep=kr_{p}(M_{p}/3)^{1/3}$ with $k<0.5$. Table \ref{tab:sims} shows the parameters of our simulations. Distances are given in units of $r_{0}=1AU$, density is given in units of $\rho_{0}=2.6\times10^{-10} g cm^{-3}$, and velocity is given in units of Keplerian speed at $1AU$,  $v_{0}=v_{k}(1AU)$. The surface density have been scaled such that the total disk mass is $0.01M_{star}$. Magnetic fields are given in units of $B_{0}=\sqrt{4\pi \rho_{0} v^{2}_{0}} $. In the cases where the planet is not on a fixed orbit (runs see Table \ref{tab:sims}), the equations of motion are integrated with a simple leap frog integrator.
For the calculation of the torque, we include the entire disk in the integration (without Hill sphere tapering, except for simulation R9). The components of the torque vector in cartesian coordinates are given by
\begin{equation}
\Gamma_{i} = GM_{p} \int \rho(\mathbf{r}) \frac{(\mathbf{r_{p}}\times\mathbf{r})_{i}}{(|\mathbf{r}-\mathbf{r_{p}}|^{2} + \ep^{2})^{3/2}}dV,
\end{equation}
where $i\in\{x,y,z\}$ is any of the three cartesian indices and $(\mathbf{r_{p}}\times\mathbf{r})_{i}=(\mathbf{r_{p}}\times\mathbf{r})\cdot \hat{e}_{i}$, with $\hat{e}_{i}$ being the cartesian unit vectors.  Of course, for studying migration, we are mostly interested in the $z$ component of the torque vector. Specific torques are given in units of $v_{k}^{2}(1AU)$.

\subsection{Magnetic Field Configuration}
\begin{figure}[th]
  \epsscale{0.6}
  \plotone{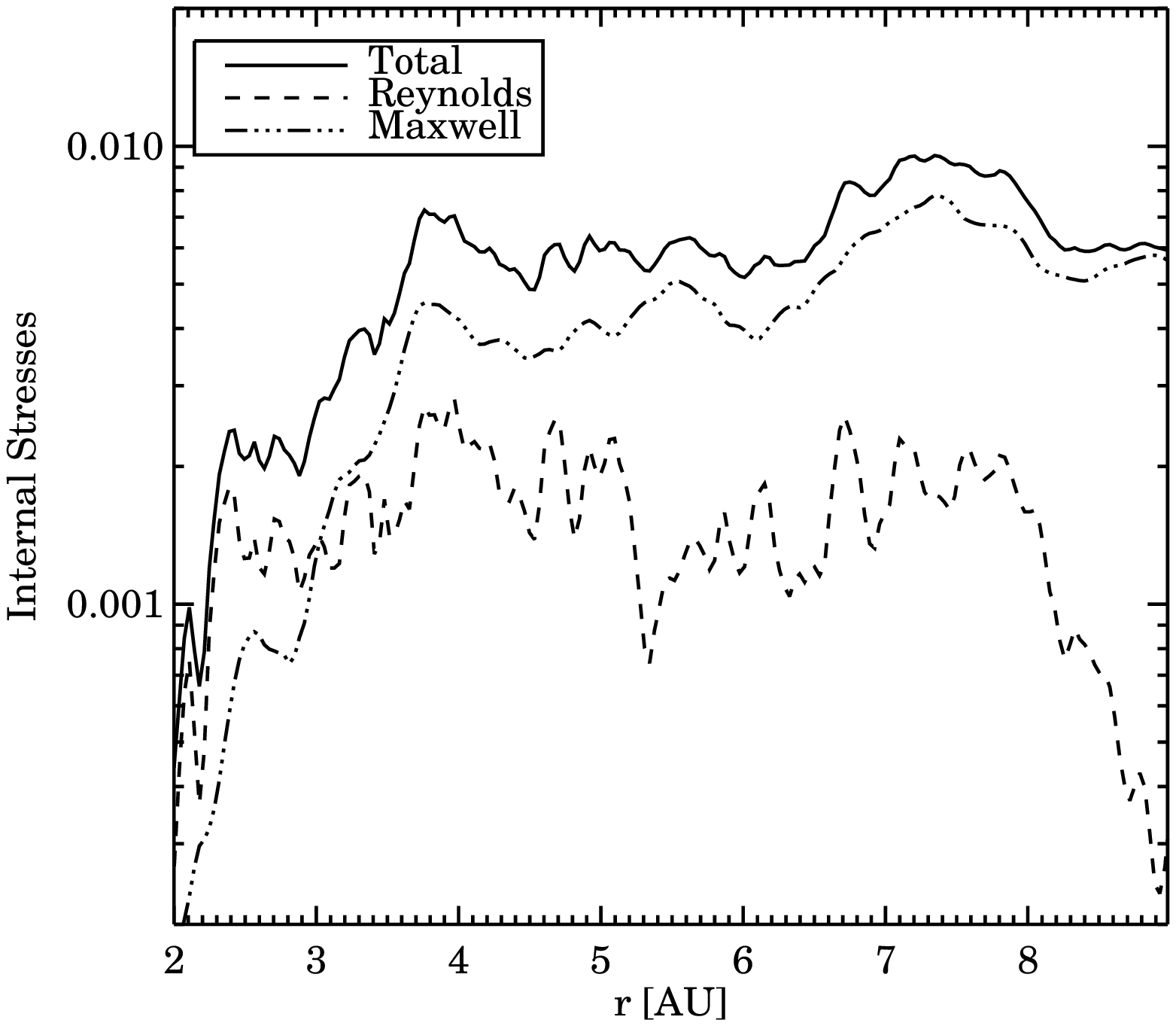}
  \caption{Initial radial distribution of the time, azimuthally and vertically averaged stress parameter (before the addition of the potential of the planet). The dashed and dashed-dot lines show the Reynolds $T_{Rey}$ and Maxwell $T_{Max}$ stresses respectively, normalized by the initial pressure. The solid line shows the total effective $\alpha$ parameter.}\label{fig:stress0}
\end{figure}

\begin{figure}[th]
  \epsscale{0.9}
  \plotone{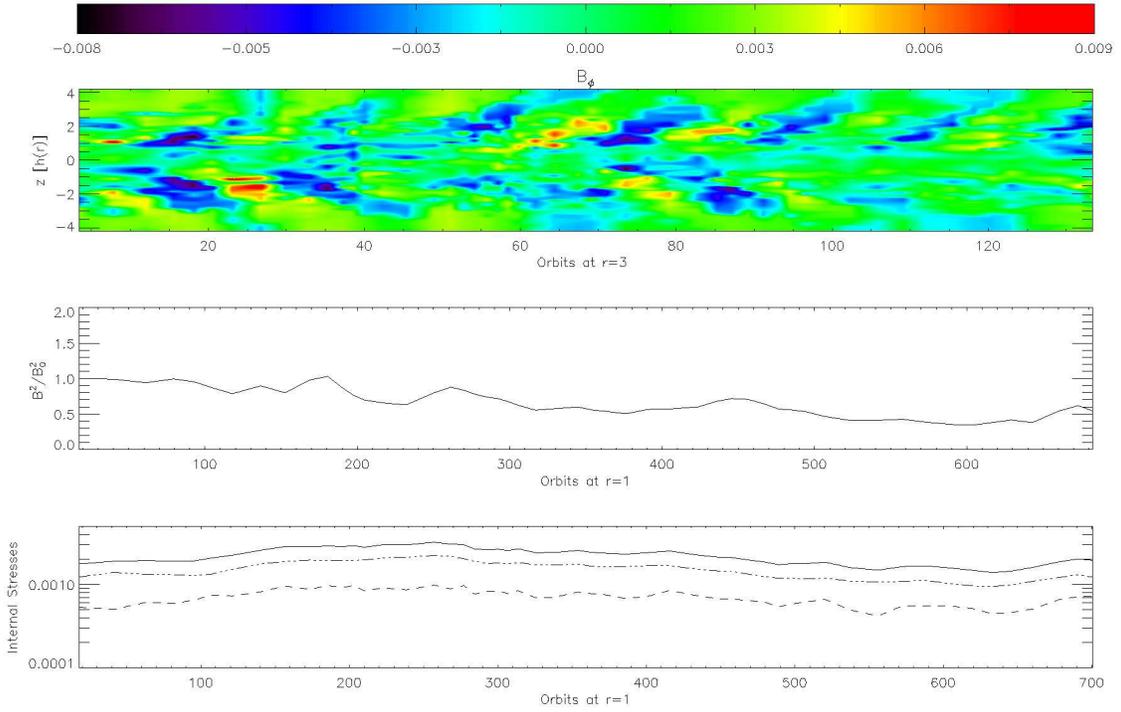}
  \caption{Time evolution of $B_{\phi}$, $B^{2}/B^{2}_{0}$ and $\alpha$ for a run without a planet. The dashed and dashed-dotted lines show the Reynolds $T_{Rey}$ and Maxwell $T_{Max}$ stresses respectively, normalized to the initial pressure. The solid line shows the total effective $\alpha$ parameter.}\label{fig:bfields}
\end{figure}

Before introducing the planet in the simulations, a weak toroidal magnetic field is imposed on the disk given by
\begin{equation}
(B_{r},B_{\theta},B_{\phi})=(0,0,2p/25),
\end{equation}
where $p$ is the initial thermal pressure. This gives an initial azimuthal field with constant plasma beta $\beta=25$. The field is imposed in a subset of the full computational domain given by $2<r<9$ and $\pi/2 - 0.07 < \theta < \pi/2 + 0.07$. The simulation is then followed until turbulence generated by the MRI has reached a saturated state. After this stage, we reset the density to the initial condition. This is the initial state in which the potential of the planet is incorporated and where all our runs start. The azimuthally, vertically and time averaged value of the effective $\alpha$ parameter is shown in Figure \ref{fig:stress0}. However, the $\alpha$ stress is not constant throughout the vertical dimension. The upper layers of the disk are the most active. Figure \ref{fig:bfields} shows the time evolution of $B_{\phi}$, $B^{2}/B^{2}_{0}$ and $\alpha$ for simulation R0 (see Table \ref{tab:sims}) that does not include a planet. The top figure shows the characteristic butterfly diagram for the azimuthal component of the magnetic field in a turbulent stratified disk. \footnote{A more complete description of the type of model used in this paper and a detailed analysis of the MRI, magnetic fields and turbulent spectra can be found in \citet{2011arXiv1104.4565F}.}

\subsubsection{Zonal flows and pressure bumps}\label{zonalflows}
\begin{figure}[h]
  \epsscale{0.7}
  \plotone{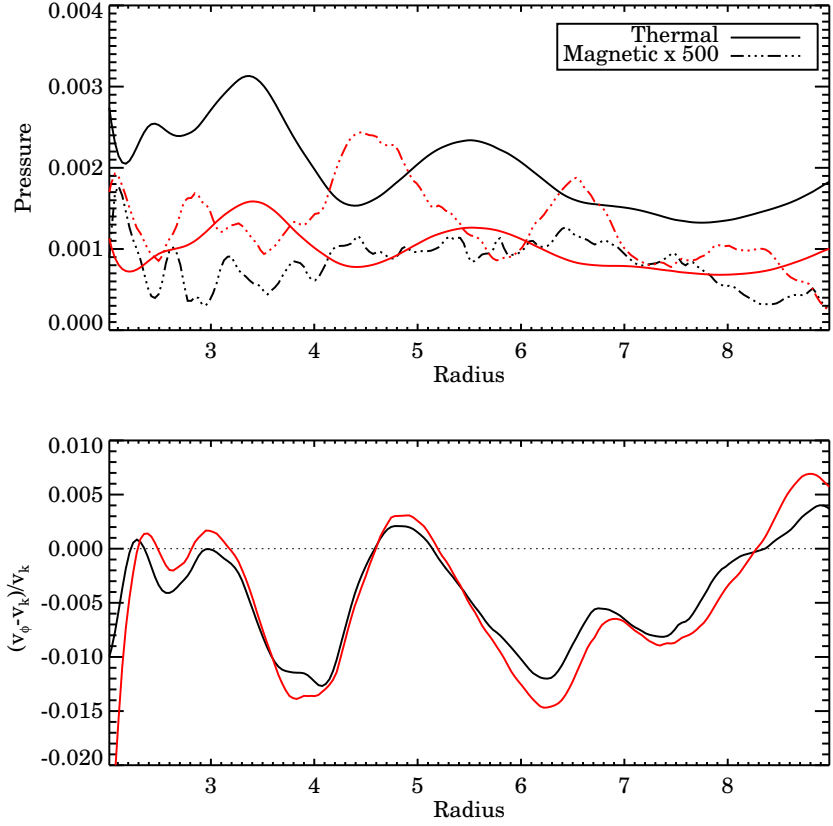}
  \caption{\textsl{Top figure:} Time-averaged thermal and magnetic pressure in the mid-plane (black line) and one scale height above the mid-plane (red line). The profiles have been normalized to take out the radial variation. \textsl{Bottom figure:} Azimuthal velocity perturbation (with respect to Keplerian speed) in the mid-plane (black line) and one scale height above the mid-plane (red line). This is from a simulation with no planet included. }\label{fig:pryvel_mhd}
\end{figure}

The time-averaged thermal and magnetic pressure and the perturbed (with respect to Keplerian) azimuthal velocity are plotted in Figure \ref{fig:pryvel_mhd} for run R0. We plot these quantities in the mid-plane of the disk, and one scale height above the mid-plane, for a simulation without a planet (or, equivalently, a massless planet). The radial gradient of the pressure has been removed and the pressure is averaged in the azimuthal direction. As expected of zonal flows, we see pressure bumps that correlate with bumps in perturbed azimuthal velocity, only phase shifted by one quarter of a period \citep{2009ApJ...697.1269J}. Bumps in thermal pressure correlate with drops in magnetic pressure, a behavior that is seen more clearly above the mid-plane, since the MRI is not resolved in the mid-plane of the disk. Notice also that in the velocity peaks, the azimuthal velocity exceeds the Keplerian value at some radial locations. These structure in the pressure and the velocity lived for the entire duration of our simulations, around 1000 inner orbits. These "zonal flows" result from an inverse cascade of kinetic energy, e.g. a transport of energy from the MRI unstable medium scales, to the largest scales, which is very typical in accretion disks simulations (see for instance \citet{2010A&A...515A..70D} and \citet{2008A&A...479..883L}).

\section{RESULTS}\label{results}

\subsection{Disk torques and migration}
Table \ref{tab:sims} summarizes the computational time in local planet orbits for each of the simulations. The torque was calculated by taking into account the entire disk and its value was saved at every time step. We calculated the cumulative average specific torque as
\begin{equation}
\Gamma_{n} = \frac{1}{T_{n}}\Sigma_{k=1}^{n}\Gamma_{k}\Delta t_{k},
\end{equation}
where $\Gamma_{n}$ is the cumulative average torque up to timestep $n$ and $T_{n}$ is the total time until timestep $n$.

\subsubsection{Low Mass Planets ($q=5\times10^{-6}$ and $q=1\times10^{-5}$)}
Figure \ref{fig:rho} summarizes the density structure of simulations R2, R5 and R9. Runs R1 ($q=5\times10^{-6}$), R2 and R3 ($q=10^{-5}$) shows no significant perturbation of the density by the planet, and no spiral arms are seen. The turbulent perturbations dominate in this case.
Figure \ref{fig:cum_tor_q4} shows the torque cumulative average torque as a function of local orbits for run R1. Figure \ref{fig:cum_tor_q3} shows the torque for simulations R2 and R3. 
The fluctuations in the torque created by the perturbations in the density can, in both cases, be larger than the mean torque expected for standard Type I migration in a laminar disk. Comparing the torque for the planet at different positions in the disk, we see that the local (in time) evolution depends on the location of the planet. Random variations in the torque can be an order of magnitude larger than torques coming from the Lindblad resonances, in addition to the possibility that the spiral waves excited at Lindblad resonances are partly or totally suppressed by density fluctuations coming from the turbulence, such that the magnitude of this torque can be reduced. For the low-mass planet simulations, we find no convergence of the torque on timescale of the runs. For a run with a massless planet orbitting at $r_{p}=3.3$, a gaussian fit of the time distribution of the torque gives a standard deviation of $\sigma \approx 1.5e-5$. We also calculate the auto-correlation function of the torque and take the correlation time to be given by $\tau_{c}=\int_{0}^{t_{max}}ACF(\tau)d\tau$. This gives $\tau_{c} \approx 2$ local orbits, while the first and second zero crossing of the torque ACF occur at $0.2$ and $0.8$ local orbits. This is in agreement with results by \citet{2005A&A...443.1067N} and \citet{2009A&A...496..597F} and with estimates used by \citet{2010ApJ...709..759B} that are based in previous MHD simulations of turbulent low-mass planet migration. We also calculate the power spectrum of the mid-plane density (see Eq. 3 in \citet{2010ApJ...709..759B}). This is shown in Figure \ref{fig:rho_pow_spec} and we compare our results from MHD simulations to Figure 1 of \citet{2010ApJ...709..759B}, where the spectrum in the result of the forcing model for the turbulence with $\alpha \sim 10^{-3}$. This comparison is valuable, since ultimately, a more complete parameter study migration and turbulence will have to be studied in models with forced turbulence. For this $\alpha$ value, in our simulations we find that the larger scales carry more power than in the random forcing model, while the two spectrum agree for smaller scales for the case that includes the modes with $m>6$. The higher power at larger scales can result from the higher compressibility of stratified disks, especially at large scales, as a vertically stratified disk can respond to compression with vertical expansion. However, the overall shape of the spectrum of the MHD simulation agrees better with the HD simulation without the $m>6$ modes included.  Therefore with the proper scaling of the amplitude of the turbulence, and a cutoff of these modes, these simulations could reproduce the MHD spectrum. Another possibility is that the power at the small scales in the MHD simulation is lower due to lacking resolution at these scales. Ultimately, there needs to be a physical motivation for the cutoff of the turbulent forcing potential after the first few modes, if this is indeed the model that better matches the global MHD simulations.

\subsubsection{Intermediate Mass Planets ($q=5\times10^{-5}$, $q=1\times10^{-4}$ and $q=2\times10^{-4}$)}

Figure \ref{fig:rho} shows the log density for run R5.  For this simulation ($q=10^{-4}$), spiral arms are visible and their amplitude is comparable (or larger closer to the planet) to that of the perturbations generated by the turbulence. Figure \ref{fig:cum_tor_q6} shows the cumulative average specific torque for run R4 and Figure \ref{fig:cum_tor_q2} shows the torque for simulations R5 and R6. We see that in these three simulations there is an initial stage where the torque is negative
followed by a reversal of the direction of the migration where the torque becomes positive and takes a defined value for the rest of the simulation. This happens at different times when we compare two different positions of the planet in the disk (runs R5 and R6). Instead of a random walk variation in semi major axis superimposed on smooth inwards migration, we find that planets of around 30 Earth masses undergo systematic outward migration. This outward migration is sustained for the total duration of the simulation. The simulation times for runs R5 and R6 are around 600 to 1000 orbits at the inner boundary of the disk (1AU). During this time, the density profile in the disk can evolve significantly from the initial state, and although the surface density profile can still be fitted by the initial profile ($\Sigma \propto r^{-1/2}$), there can be changes in the local profile at the position of the planet and accumulation of mass at the disk inside the planet's orbit due to turbulent stresses. However, for both simulations, the torque is reversed before there is a significant accumulation of mass at the inner boundary and it converges to a constant value for the remaining simulation time. 
The torque for run R8 is shown in Figure \ref{fig:cum_tor_q5}. In this run the planet mass ($q=2\times10^{-4}$) is now able to modify the density profile around its orbit, and opens a partial gap, which affects the convergence of the torque. We don't find convergence for the simulation time, but there is still a tendency for outwards migration.

Unlike the simulations for the small-mass planets (R1, R2 and R3), for the simulations R4, R5 and R6, the hill radius of the planet and the horseshoe region are resolved (by approximately 4, 4 and 7 grid cells per half width respectively). In this case, the component of the torque originating from the horseshoe region can dominate if there is a mechanism for keeping the corotation torque unsaturated and the local density profile differs from the global profile, possibly increasing outwards, such that the corotation torque can be larger than the Lindblad torque, making the total torque positive. There are special locations in the disk where is possible for the surface density to increase outwards, due to the appearance of zonal flows, as seen in Section \ref{zonalflows}. \\

Comparing the torque values of the simulations with analytical estimates by \citet{2009MNRAS.394.2283P} or \citet{2006ApJ...652..730M} is not straightforward. First, the undergoing evolution of the disk can make the surface density profile at the position of the planet and the effective stress resulting from turbulence vary in time, therefore one torque estimate does not apply at all times. On the other hand, the value of the horseshoe drag is very sensitive to the structure of the horseshoe region, and the estimate used in the analytical calculations is based on a 2D model of the flow around the planet. In our case, the horseshoe region is distorted, making the half-width difficult to define. Additionally, the inclusion of magnetic fields can introduce new magnetic resonances that affect the total torque, as seen by \citet{2003MNRAS.341.1157T} and \citet{2005MNRAS.363..943F} for a uniform non-turbulent field. For the sake of the comparison and simplicity, we discard this type of contribution.
We attempt a comparison with the analytical estimates for the torque, including a contribution of the horseshoe drag. We take the total torque to be composed of the Lindblad torque in a 3D locally isothermal disk \citep{2002ApJ...565.1257T}

\begin{equation}
\Gamma_{Lind} = -(2.340 - 0.099a + 0.418b)\left(\frac{q}{h_{p}}\right)^{2}\Sigma_{p} r_{p}^{4} \Omega_{p}^{2},
\end{equation}  
plus the fully unsaturated non-linear horseshoe drag\footnote{We take the expression for an isothermal disk, in the zero gravitational softening limit.} 
\citep{2009MNRAS.394.2283P}
 
\begin{equation}
\Gamma_{HS} = \frac{3}{4}\left(\frac{3}{2}-a\right)x_{s}^{4}\Sigma_{p} r_{p}^{4}\Omega_{p}^{2}.
\end{equation}  

Here $\Omega_{p}$ is the angular frequency of the planet and $\Sigma_{p}$ is the surface density at the position of the planet. The cumulative average torque at the end of the simulation for run R5 is $2.0\times10^{-5}$ and we take the half width of the horseshoe region to be $x_{s}=0.25$, as is measured in our simulations (calculated from the analytical expression, $x_{s}=0.24$). Assuming the global surface density profile $\Sigma\propto r^{-d}$, with $d=0.5$, will always give a negative torque. However, the torque always becomes positive for $d=0.3$ and matches the simulation value for $d=-1.5$, which is comparable to the local profile observed in the simulations (see middle plot in Figure \ref{fig:surf_dist_evol}). We should also note that already a "close-to-flat" profile can significantly reduce the negative torque or change the sign of the torque. For simulation R6, the cumulative average torque at the end of the simulation is $3.7\times10^{-6} $ and the measured half width of the horseshoe region is $x_{s}=0.16$. In this case, a local profile of $d=0.1$ is enough to obtain a positive torque, while a local profile of $d=-0.4$ matches the value of the torque obtained in the simulation. However for this simulation, we only observe a flatter profile (see Figure \ref{fig:surf_dist_evol}). This discrepancy can be due to the fact that these values are very sensitive to the value of $x_{s}$, since the horseshoe drag scales as $x_{s}^{4}$, and we stress that the streamlines can be very distorted, therefore making the estimation of $x_{s}$ difficult. This is a critical parameter, and we find that an increase of $ 1\% $ to $5\%$ in the simulation value of $x_{s}$ with respect to the analytical estimation is enough to reproduce the observed positive torques. Therefore, if one assumes that the observed torque is composed of the wave torque plus the corotation torque and neglects any additional effect, we see that the corotation torque is crucial and able to cancel out or overcome the negative Lindblad contribution for standard disk parameters. 
\\
To further test the effect of the local density profile, we performed a simulation with $q=10^{-4}$, for the planet located at $r_{p}=4.0$ (run R7, Figure \ref{fig:cum_tor_q2_r3}), initially at the right side of a pressure bump (where pressure and density decrease with radius). In this case, the cumulative average torque does not clearly converge, and we do not see systematic outwards migration, as the cumulative average torque approaches zero. However there is still a significant reduction of the torque as compared to the Type I Lindblad torque, which cannot be explained only in terms of a locally decreasing radial density profile. This result suggest that even in the absence of a pressure bump, inwards migration can be significantly slowed down for this planet mass. We note also that we used the expression for the horseshoe drag valid for an isothermal disk, so that there is an additional contribution due to the locally isothermal profile that we did not take into account.  
\\
To see if the transport of mass in the disk is enough to sustain the unsaturated torque, we take the expression for the minimal $\alpha$ to mantain the unsaturated corotation torque \citep{2006ApJ...652..730M}
\begin{equation}
\alpha_{m} = 0.035q^{3/2}h^{-7/2},
\end{equation}   
we obtain $\alpha_{m} = 0.0003$  for $q=10^{-4}$, which is always smaller than what we observe in our simulations\footnote{However, this expression for $\alpha_{m}$ is derived using a 2D model of the HS region, which determines the viscous crossing time across the region, the libration time and the U-turn time.} (comparing with the volume average $\alpha$). For run R6 we also observed that in comparison to a purely HD laminar run, in which the planet is able to open a partial gap in the disk, the gap in this case is less deep that in the HD case, and also wider, compared to the narrower gap seen in the laminar simulation. For run R5, there was no gap opening neither in the laminar nor the turbulent runs as the gap opening criterion is not satisfied. We also find that the stresses in the disk are affected by the presence of the planet; the volume averaged stress decreases as the mass of the planet is increased (see top and middle plots in Figure \ref{fig:stresses_qs}), which might be a result of numerical dissipation due to the limited resolution.

\subsubsection{Large Mass Planet ($q=10^{-3}$)}
\begin{figure}[th]
  \epsscale{0.9}
  \plotone{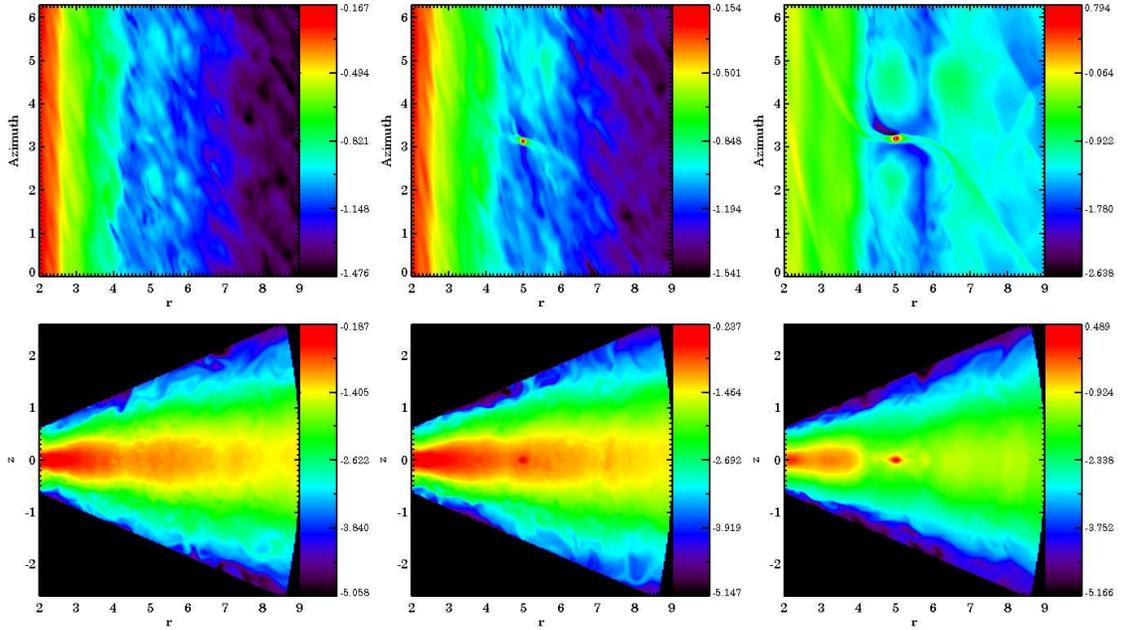}
  \caption{Logarithm of the disk density in the mid plane (top row) and in an azimuthal cut at the position of the planet (bottom row) for runs R2 (left, $q=10^{-5}$), R5 (middle, $q=10^{-4}$) and R9 (right, $q=10^{-3}$).}\label{fig:rho}
\end{figure}

\begin{figure}[th]
  \epsscale{0.6}
  \plotone{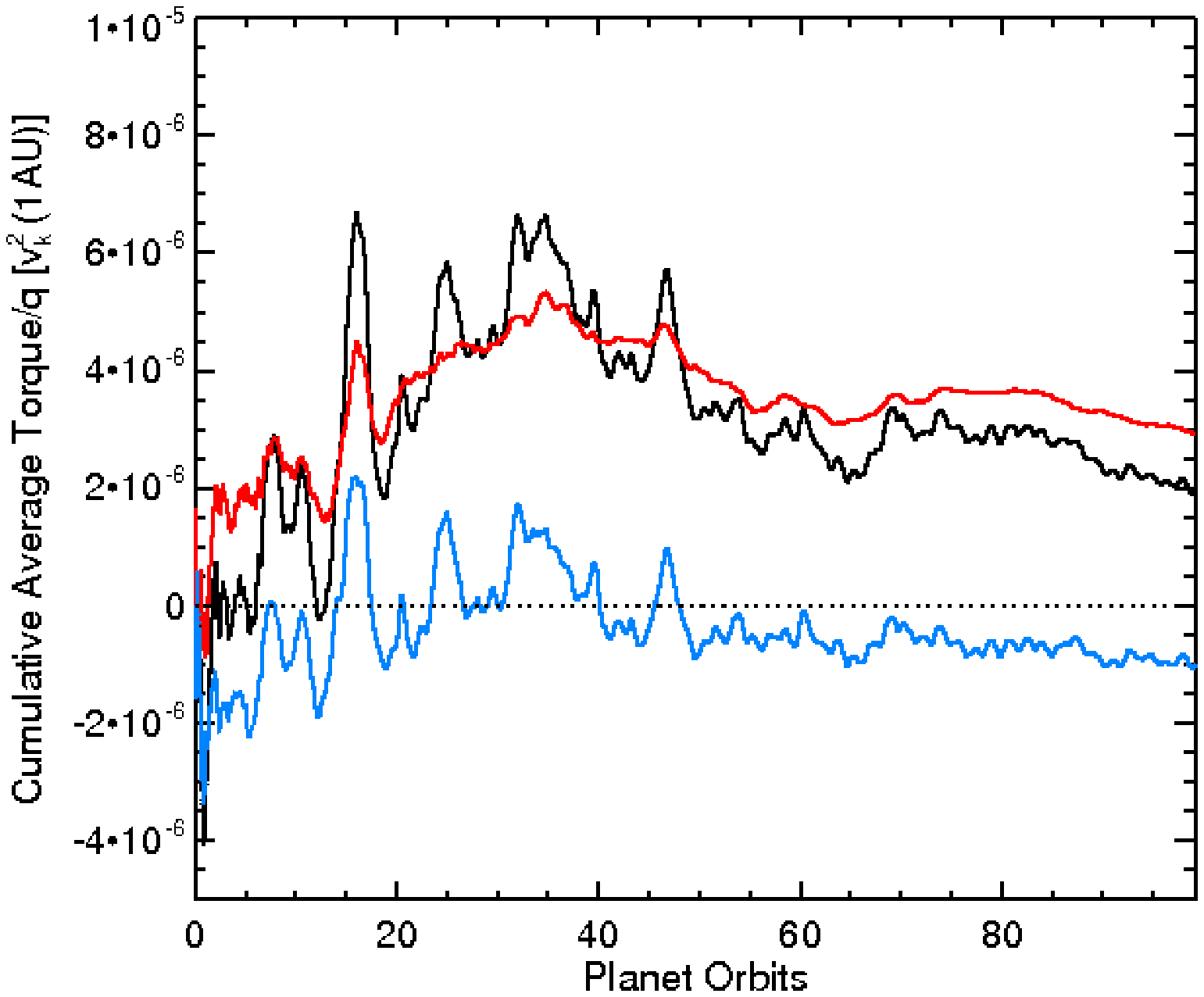}
  \caption{Cumulative average torque for run R1 for $q=5\times10^{-6}$. The red and blue lines show the torque exerted by the inner and outer disk respectively.}\label{fig:cum_tor_q4}
\end{figure}

\begin{figure}[th]
  \epsscale{0.9}
  \plotone{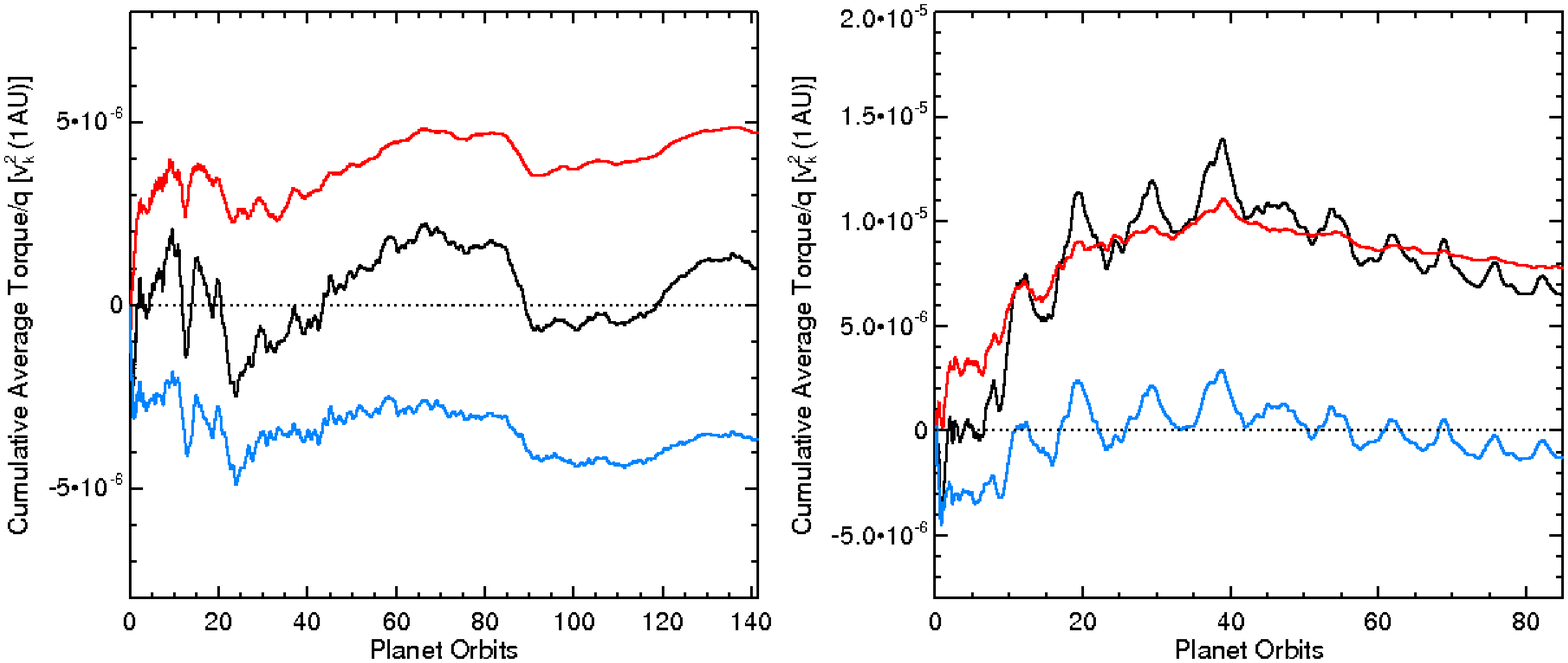}
  \caption{Cumulative average torque for runs R2 and R3, for $q=10^{-5}$,where the planet is located at $r_{p}=3.3$ and $r_{p}=5.0$ respectively. The red and blue lines show the torque exerted by the inner and outer disk, respectively. }\label{fig:cum_tor_q3}
\end{figure}

 The density structure and spiral arms induced by the Jupiter mass planet in run R9 ($q=10^{-3}$) dominate over the turbulent perturbations and influences the entire disk. Figure \ref{fig:cum_tor_q1} shows the cumulative average torque. We have excluded the torques coming from the Hill sphere for the calculation of the torque, to exclude material bound to the planet which is not properly simulated at this resolution or without including other relevant effects. For these simulations, initially we see the same trend as for runs R5 and R6, where the torque becomes positive, however this is not sustained for this planet mass as torques coming from the corotation region are suppressed due to gap opening (see bottom plot in Figure \ref{fig:surf_dist_evol} and the right panels in Figure \ref{fig:rho}). Additionally, the planet modifies the stresses in the disk, and therefore, the accretion behavior, as can be seen in the bottom plot in Figure \ref{fig:stresses_qs}, that shows the evolution of the stresses for run R9. In this case, the $\alpha$ stress is progressively suppressed and the Reynolds stress dominates the total stress. The reduction of the Maxwell stress is seen mostly in the part of the disk inside the planet's orbit and in the gap region (see Figure \ref{fig:stresses_r_q1}). Is possible that the magnatic stress is suppressed in the gap region due to the modified azimuthal velocity near the planet, which can suppress the MRI locally. Notice that for this run, the density has a more laminar appearance (see Figure \ref{fig:rho}), consistent with a reduction of the stress due to the presence of the planet. This could also be a numerical effect that appears at this resolution, so further studies at higher resolution are needed.
\\ 
In Figure \ref{fig:gap_q1_comparison} we compare the gap opened by the Jupiter mass planet in a magnetized disk (run R9) with an equivalent HD 3D simulation with $\alpha$ viscosity where $\alpha = 2\times10^{-3}$ (run R10) and stratification. The time of the snapshots is 100 local orbits. The gap for the hydro case is narrower and slightly deeper than the gap formed in the magnetized turbulent disk. However the gap is not completely cleaned after this time. We observed the same characteristics for lower-mass planets that open only a partial gap in the disk.\\
\citet{2003ApJ...589..543W} studied a similar case of gap opening, but in a unstratified MHD-turbulent disk. In agreement with our results, they found a wider gap when the disk is turbulent, and larger transport of mass from the outer to the inner disk (see our Figure \ref{fig:gap_q1_comparison}). In contrast to our findings, they find a deeper gap in the hydro case. This can be due to a different treatment of the gap opening criteria, since the planet mass in their calculations does not satisfy the viscous criterion. In terms of the reduction of the stresses around the planet, we find agreement with their results. \\
\citet{2003MNRAS.339..993N} studied gap opening by a giant planet in an MHD-turbulent unstratified disk and compared their results with 2D simulations with an $\alpha$ viscosity. They found that a run with an equivalent $\alpha$ stress to the turbulent run produced a shallower gap. This is different to our own results where we found the same gap depth. However, we should note they only studied how the turbulence affects an already formed gap and did not observe the depletion of the outer disk. The difference to our simulation results could also be due to our choice of $\alpha$ in the hydro run. We choose an $\alpha$ matching the global average of the turbulent runs, but this is a quantity that varies vertically. In a non-stratified disk simulation, this averaging is not necessary.

\section{DISCUSSION AND CONCLUSIONS}\label{conc}

For simulations R1, R2 and R3, where $q=10^{-5}$, during the simulated time, migration was dominated by random fluctuations in the torque, that can be orders of magnitude larger that what is expected for the value of the Lindblad or corotation torques for this planet mass. This is in agreement with simulations by \citet{2005A&A...443.1067N} of migration of low-mass protoplanets in cylindrical disk models, where stratification is neglected. 
It is unclear if after long term averaging ($\sim 1000$ orbits), the fluctuations will average out to zero while some component of the systematic torque will remain. Such a calculation is currently too expensive. It will also be difficult to get a steady state without special prescriptions for correcting the density, due to the accretion evolution of the disk, in addition to the decrease in $\alpha$ stress for long simulation times due to the limited resolution. Another interesting point for further studies is to investigate this type of migration with enough resolution to resolve the corotation region, to see the impact of the corotation torque in these cases. However, even if this torque is present and well resolved, its magnitude would still be small compared to the amplitude of the fluctuations, since ultimately the torque depends strongly on the width of the corotation region, which approaches zero as the planet mass approaches zero.   \\

As the planet mass is increased by one order of magnitude to $q=10^{-4}$, the hill radius is now properly resolved and the systematic torque is now large enough to dominate over the random component of the torque. Outwards migration in a locally isothermal disk can occur due to the viscosity unsaturating the torque coming from the corotation region (where the viscous timescale across the horseshoe region is smaller than the libration timescale), as was found by \citet{2006ApJ...652..730M} for planets in the intermediate-mass range. Specifically, a planet with mass ratio $q\approx10^{-4}$ in a disk with $h=0.06$ with a flat surface density profile ($d=0$) and an $\alpha$ viscosity was found to be the critical mass for which the offset from linear theory was the largest. Additionally, a planet mass of $q=10^{-4}$ is within the range of masses for which reversal of migration occurs, if one extrapolates their results to a disk with $h=0.07$. However, this is the first study that observes the effect of the unsaturated corotation torque due to the accretion of mass in the disk provided directly by turbulence that has been self-generated by the MRI.
 For runs R4, R5 and R6, the planet is in locations in the disk where the local surface density profile is either close to flat or increasing outwards, due to the pressure bumps seen in Section \ref{zonalflows}. This can make the contribution of the corotation torque dominate over the Lindblad torque, which is usually unexpected in an $\alpha$ disk, since for realistic density profiles, the Lindblad torque will dominate. This is also consistent with the torque in run R6 being initially negative, since at the beginning of the simulation the local profile is decreasing outwards, but getting shallower as time increases, eventually reaching the point where the torque reverses. For run R5, the slope is almost immediately increasing outwards due to the evolution of the disk, which makes the torque positive from the beginning of the simulations. We can only roughly compare our numerical results with analytical estimates, as was done in the previous section, for the reasons described already there. Also in comparing with previous estimates, we also discarded any possible additional contributions to the torque that might arise because of the turbulent magnetic fields. The detailed structure of the horseshoe region in the presence of turbulence and stratification deserves further study. Our results are summarized on Figure \ref{fig:results_comp}, where the torque dependence on planet mass is shown. For each simulation, we plot the last value of the cumulative average torque. Note however that only for part of the simulations the torque converges to a well defined value. It is possible to see a trend of the torque to reverse, corresponding to the addition of the contribution of the fully unsaturated horseshoe drag ($\Gamma_{tot} = \Gamma_{Lind}+\Gamma_{HS}$). For the plot we assumed values for the width of the horseshoe region that are $5\%$ larger than the analytical estimate given by \citet{2009MNRAS.394.2283P} and we use the value of the local surface density profile for the calculation of the torque. We see that the trend breaks down already for $q=2\times10^{-4}$, where gap opening starts to become important and there is a transition into the Type II regime. Error bars represent the standard deviation of the time distribution of the torque. Note that since the raw torque is a highly oscillating quantity, the standard deviation does not match directly to the amplitude of the turbulent fluctuations, especially in the high-mass planet cases. For run 10, the standard deviation was found to be only 20\% lower that in the turbulent run R9. In the Type II range, we plot the torque corresponding to the viscous timescale of the disk, taking $\alpha=3\times10^{-3}$. We find reasonable agreement with our simulation, taking into account the short simulation time, and that the value of the torque is still decreasing in the simulation after 100 orbits. Additionally we use the value of the initial, volume averaged $\alpha$, while the mid-plane value is smaller. 

The question remains about the long term behavior of the torque, and whether this is only a transient behavior lasting for the first few hundred orbits (assuming the same local surface density profile), afterwards saturating and returning to standard negative Type I values. This is still a transient behavior in the sense that the planet can migrate out of the part of the disk where the local profile allows for outwards migration and enter a region where migration proceeds inwards again. Additionally it is limited by the lifetime of the pressure bumps, which we weren't able to determine. We observe a stable pressure bump through the duration of our simulations.  If there are other mechanisms such as the ones discussed in \citet{2006ApJ...642..478M} that produces this type of locally increasing outwards density profile, then, in the presence of turbulence, these density bumps can also act as a protoplanet trap and halt, slow down or reverse inwards migration. \citet{2010A&A...515A..70D} performed non-ideal MHD simulations of accretion disks with spatially varying resistivity. They also find zonal flows/pressure bumps not only at the snow line, e.g. a region with a jump in resistivity, but also inside the more active region. They already suggest that small planets should get trapped at those local pressure maxima (see also \citet{2007ApJ...664L..55K}).

\section{SUMMARY}

We studied the migration of planets under the influence of turbulence that is a result of the magneto-rotational instability. We find that, under the right conditions, planets can undergo systematic outwards migration in a locally isothermal disk. After long term averaging, transient or long term periods of outwards migration can help the survival and influence the mass accretion history of giant planet cores of a certain mass ratio. The contribution of the unsaturated horseshoe drag and the stochastic migration of low-mass planets, which are both consequences of the turbulence, should be incorporated into planet population synthesis models in order to test the influence of this element on the produced populations of planets. On future work we plan on studying low-mass planet migration in detail using similar stratified disk models.   \\ 

Giant planets significantly decrease the magnetic stresses in the disk (mostly inside its orbit), effectively killing the turbulence, as we observe in our simulations. This is possibly a numerical effect and it will affect the accretion behavior of the disk and possibly the Type II migration rate of the giant planet. This issue deserves further study, with high resolution simulations to determine any possible effects of numerical dissipation of the magnetic fields induced by the presence of the planet. Additionally, in agreement with previous studies, we find that the gap opened by a planet in the presence of turbulence is wider than the gap produced in a quasi-laminar disk with an equivalent $\alpha$ viscosity.

\begin{figure}[th]
  \epsscale{0.6}
  \plotone{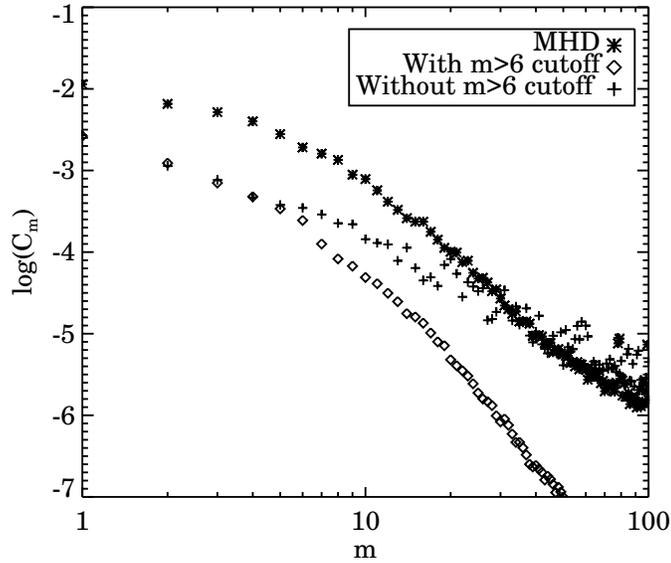}
  \caption{Power spectrum of the surface density, averaged in time and azimuthally, from MHD simulation (\emph{stars}). We compare with the power spectrum that results from the turbulent model of \citet{2010ApJ...709..759B} used in HD simulations, with (\emph{triangles}) and without (\emph{crosses}) the cutoff of the modes with $m>6$, and with effective $\alpha \sim 10^{-3}$.}\label{fig:rho_pow_spec}
\end{figure}

\begin{figure}[th]
  \epsscale{0.6}
  \plotone{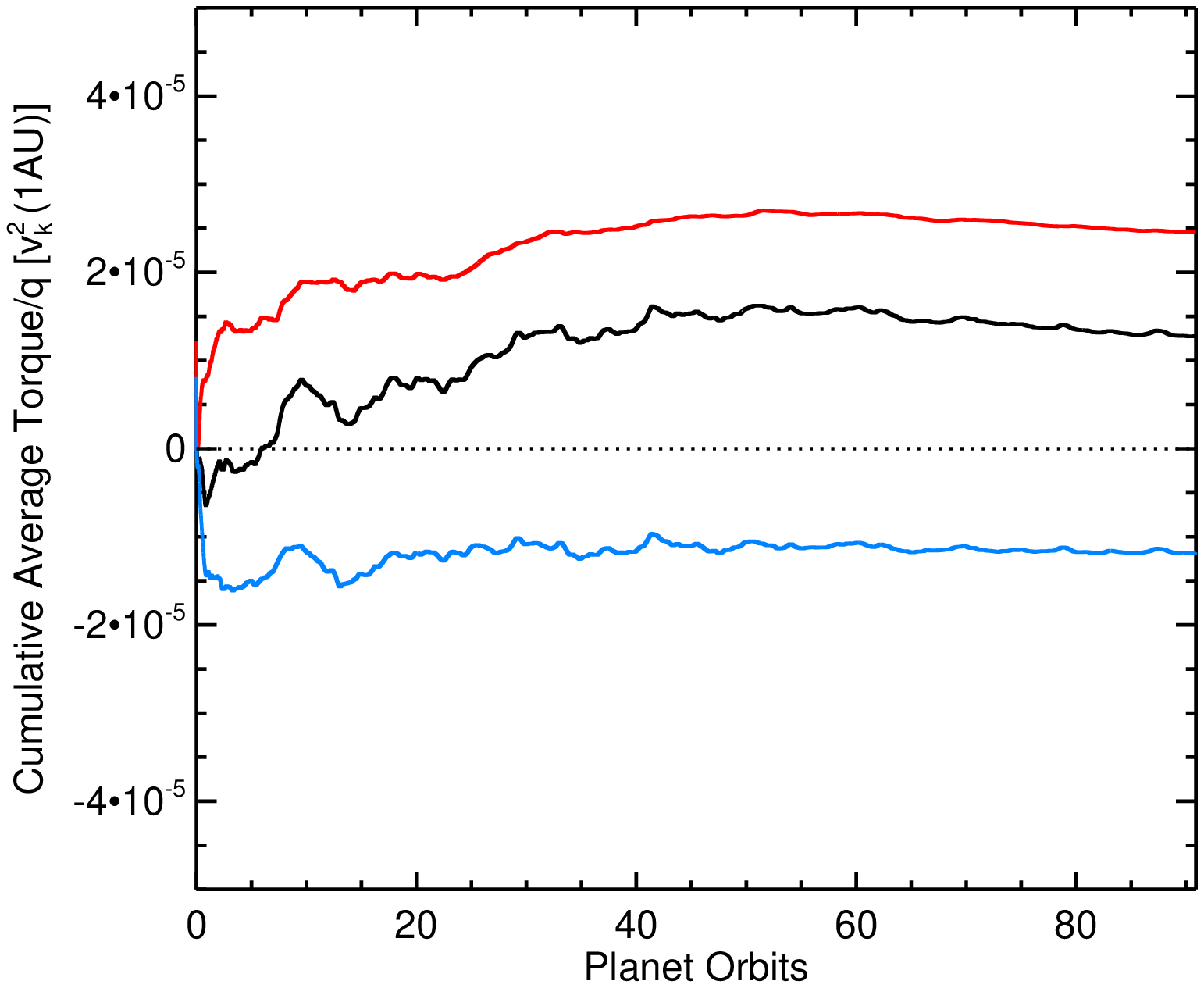}
  \caption{Cumulative average torque for run R4 for $q=5\times10^{-5}$. The red and blue lines show the torque exerted by the inner and outer disk respectively.}\label{fig:cum_tor_q6}
\end{figure}

\begin{figure}[th]
  \epsscale{0.9}
  \plotone{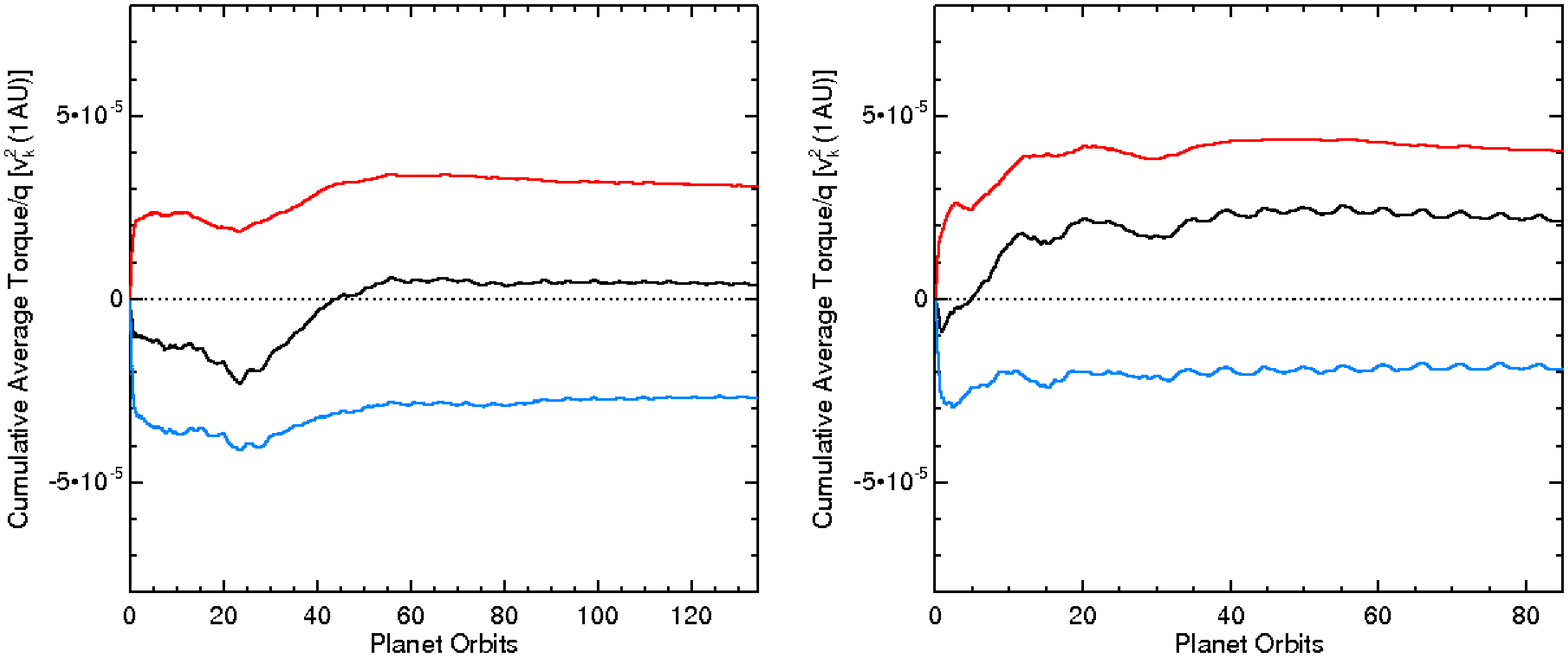}
  \caption{Cumulative average torque for runs R5 and R6, for $q=10^{-4}$, where the planet is located at $r_{p}=3.3$ and $r_{p}=5.0$ respectively. The red and blue lines show the torque exerted by the inner and outer disk, respectively.}\label{fig:cum_tor_q2}
\end{figure}

\begin{figure}[th]
  \epsscale{0.6}
  \plotone{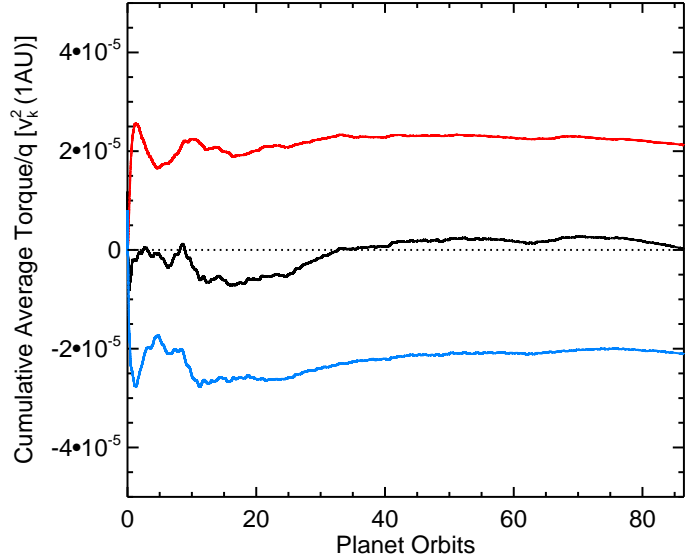}
  \caption{Cumulative average torque for run R7 for $q=10^{-4}$, for the planet located at $r_{p}=4.0$, initially at the right side of a pressure bump. The red and blue lines show the torque exerted by the inner and outer disk respectively.}\label{fig:cum_tor_q2_r3}
\end{figure}

\begin{figure}[th]
  \epsscale{0.6}
  \plotone{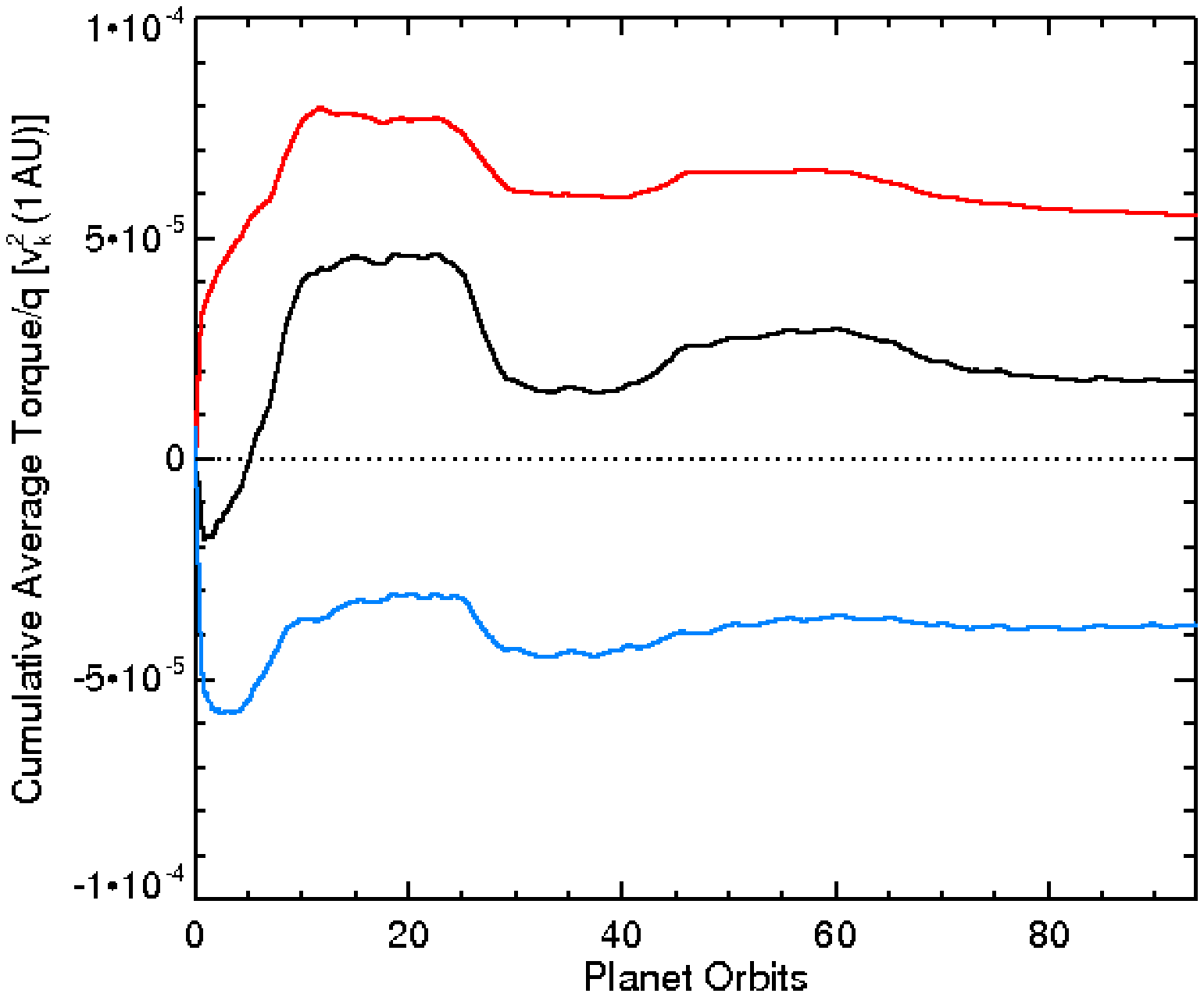}
  \caption{Cumulative average torque for run R8 for $q=2\times10^{-4}$. The red and blue lines show the torque exerted by the inner and outer disk respectively.}\label{fig:cum_tor_q5}
\end{figure}

\begin{figure}[ht]
  \epsscale{0.9}
  \plotone{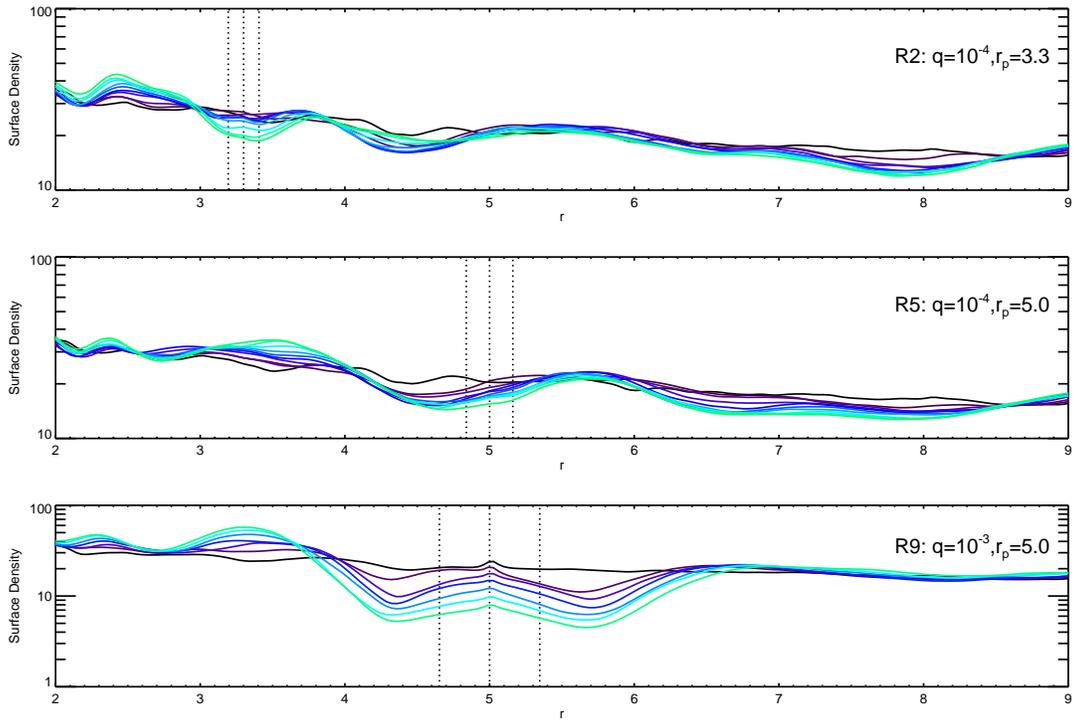}
  \caption{Surface density at different times in the simulation. Top, middle and bottom plot show the surface density for runs R2, R5 and R9 respectively. The vertical lines shows the position of the planet  and the extent of the Hill radius.}\label{fig:surf_dist_evol}
\end{figure}

\begin{figure}[ht]
  \epsscale{0.9}
  \plotone{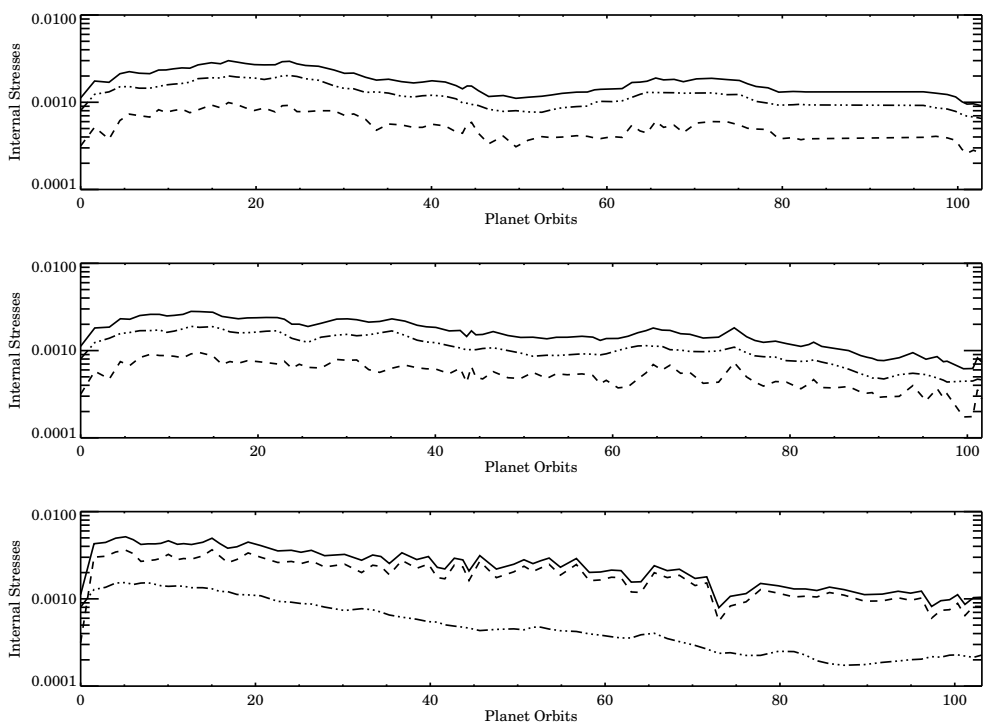}
  \caption{Time evolution of the stresses in a disk with an embedded planet. Top, middle and bottom plot show the stresses for runs R2, R5 and R9, respectively. The dashed and dashed-dotted lines show the Reynolds $T_{Rey}$ and Maxwell $T_{Max}$ stresses, respectively, normalized to the initial pressure. The solid line shows the total effective $\alpha$ parameter.}\label{fig:stresses_qs}
\end{figure}

\begin{figure}[ht]
  \epsscale{0.6}
  \plotone{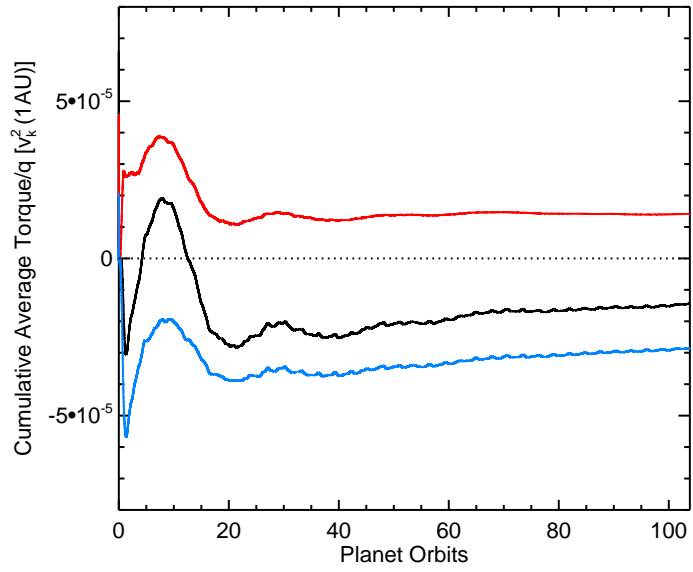}
  \caption{Cumulative average torque for run R9 for $q=10^{-3}$. The red and blue lines show the torque exerted by the inner and outer disk respectively. The torque coming from the Hill sphere has been excluded from the calculation.}\label{fig:cum_tor_q1}
\end{figure}

\begin{figure}[ht]
  \epsscale{0.6}
  \plotone{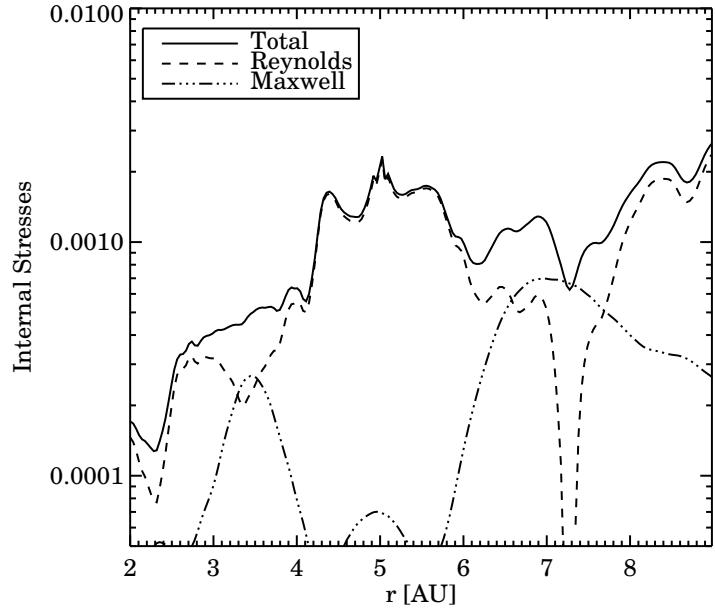}
  \caption{Radial distribution of the time, azimuthally and vertically averaged stress parameter for run R9. The dashed and dashed-dot lines show the Reynolds $T_{Rey}$ and Maxwell $T_{Max}$ stresses respectively, normalized by the initial pressure. The solid line shows the total effective $\alpha$ parameter.}\label{fig:stresses_r_q1}
\end{figure}

\begin{figure}[ht]
  \epsscale{0.6}
  \plotone{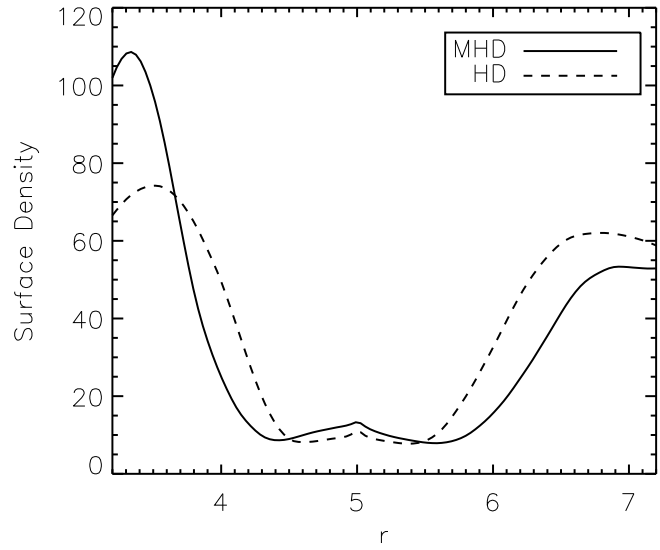}
  \caption{Gap comparison for run R9 ($q=10^{-3}$) and run 10, an equivalent HD simulation with $\alpha=2 \times 10^{-3}$.}\label{fig:gap_q1_comparison}
\end{figure}

\begin{figure}[th]
  \epsscale{0.9}
  \plotone{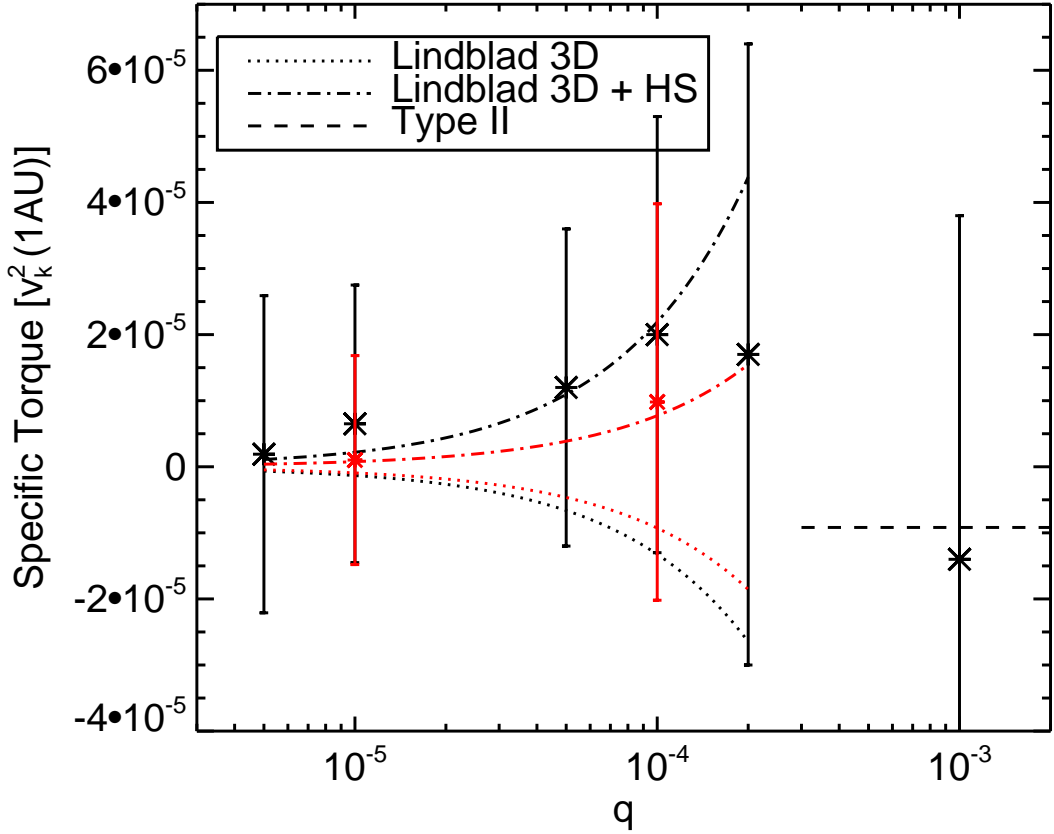}
  \caption{Specific torque as a function of $q=M_{p}/M_{s}$. The black symbols correspond to simulations R1, R2, R4, R5, R8 and R9, where the position of the planet is $r_{p} = 5.0$. The red symbols correspond to simulations R3 and R6, where the position of the planet is $r_{p}=3.3$. In the $q=5\times10^{-6}$ to $q=2\times10^{-4}$ mass range, we overplot the analytical estimates for the torque, taking into account only the Lindblad contribution $\Gamma_{tot}=\Gamma_{Lind}$ (dotted line) and both the Lindblad plus the unsaturated horseshoe drag $\Gamma_{tot}=\Gamma_{Lind}+\Gamma_{HS}$ (dash-dotted line), for both positions (i.e. local surface density profiles) of the planet, $r_{p}=5.0$ (black line) and $r_{p}=3.3$ (red line). For the analytical expressions of the torque, we take the half-width of the horseshoe region to be $5\%$ larger than its analytical estimate. The dashed line corresponds to the constant Type II migration rate, given by the viscous transport in the disk, using $\alpha=2\times10^{-3}$. Error bars represent the standard deviation of the torque time distribution. }\label{fig:results_comp}
\end{figure}

\acknowledgments
We would like to thank the support of the International Max Planck Research School (IMPRS). We also thank 
C. Baruteau for his collaboration in providing the data for the comparison in Figure \ref{fig:rho_pow_spec}. 
Simulations were performed with the IBM BlueGene/P of the Max Planck
Society and the THEO and PIA clusters of the Max-Planck-Institut
f\"{u}r Astronomie at the Rechenzentrum Garching.

\clearpage
%
%

\clearpage

\clearpage

\clearpage
%
%
\begin{deluxetable}{cccccc}
\tablecolumns{6}
\tablewidth{0pc}
\tablecaption{Simulation Parameters\label{tab:sims}}
\tablehead{
 \colhead{Name} & 
 \colhead{$q=M_{p}/M_{\star}$} & 
 \colhead{$\epsilon (r_{h})$} & 
 \colhead{$r_{p}$} & 
 \colhead{Fixed orbit} &
 \colhead{Run time (local orbits)}}
\startdata
R0 & Massless & 0.3 & 3.3 & Yes & 130\\ 
R1 & $5\times10^{-6}$ & $0.3$ & 5.0 & No & 85\\
R2 & $10^{-5}$ & $0.3$ & 5.0 & No & 140 \\
R3 & $10^{-5}$ & $0.1$ & 3.3 & Yes & 89 \\
R4 & $5\times10^{-5}$ & $0.3$ & 5.0 & No & 90 \\
R5 & $10^{-4}$ & $0.3$ & 5.0 & No & 100 \\
R6 & $10^{-4}$ & $0.1$ & 3.3 & Yes & 135 \\
R7 & $10^{-4}$ & $0.3$ & 4.0 & No & 85 \\
R8 & $2\times10^{-4}$ & $0.3$ & 5.0 & No & 95 \\
R9 & $10^{-3}$ & $0.3$ & 5.0 & No & 100 \\
R10 (HD) & $10^{-3}$ & $0.3$ & 5.0 & No & 100 \\
\enddata
\end{deluxetable}

\end{document}